\documentclass[12pt]{article}
\usepackage{graphicx}
\pdfoutput=1

\newcommand\pubnumber{}
\newcommand\pubdate{\today}

\def\institute{Departamento de F\'{i}sica Te\'{o}rica, Universidad Aut\'{o}noma de Madrid \\
Ciudad Universitaria de Cantoblanco, 28049 Madrid, Spain
\def\support{}}

\def\Title#1{\begin{center} {\Large #1 } \end{center}}
\def\Author#1{\begin{center}{ \sc #1} \end{center}}
\def\Address#1{\begin{center}{ \it #1} \end{center}}

\newcommand\pubblock{\rightline{\begin{tabular}{l} \pubnumber\\
         \pubdate  \end{tabular}}}
\newenvironment{Abstract}{\begin{quotation}  }{\end{quotation}}
\newenvironment{Presented}{\begin{quotation} \begin{center} 
             PRESENTED AT\end{center}\bigskip 
      \begin{center}\begin{large}}{\end{large}\end{center} \end{quotation}}
\def\Acknowledgements{\bigskip  \bigskip \begin{center} \begin{large}
             \bf ACKNOWLEDGEMENTS \end{large}\end{center}}

\def\beq{\begin{equation}}
\def\eeq#1{\label{#1}\end{equation}}
\def\eeqn{\end{equation}}

\def\beqa{\begin{eqnarray}}
\def\eeqa#1{\label{#1}\end{eqnarray}}
\def\eeqan{\end{eqnarray}}

\let\bar=\overbar

\def\Dslash{\not{\hbox{\kern-4pt $D$}}}
\def\dslash{\not{\hbox{\kern-2pt $\del$}}}

\def\msb{{\bar{\ssstyle M \kern -1pt S}}}

\newcommand {\pb} {\ensuremath{~\mathrm{pb}}}
\newcommand{\ifb} {\ensuremath{~\mathrm{fb}^{-1}}}

\newcommand{\GeV} {\ensuremath{~\mathrm{GeV}}}
\newcommand{\TeV} {\ensuremath{~\mathrm{TeV}}}

\newcommand{\VIIItev} {{\ensuremath{\sqrt{s}= 8\TeV}}}
\newcommand{\XIIItev} {{\ensuremath{\sqrt{s}=13\TeV}}}

\newcommand{\pp}      {\ensuremath{pp}}
\newcommand{\ttbar}   {\ensuremath{t\bar{t}}}
\newcommand{\w}       {\ensuremath{\mathrm{W}}}
\newcommand{\Z}       {\ensuremath{\mathrm{Z}}}
\newcommand{\Wjets}   {\ensuremath{\w+\mathrm{jets}}}
\newcommand{\Zprime}  {\ensuremath{\Z^{\prime}}}

\newcommand{\pt}      {\ensuremath{p_{T}}}

\newcommand{\mttbar}  {\ensuremath{M_{\ttbar}}}

\newcommand{\xtagged} [1] {\ensuremath{\textrm{#1-tagged}}}
\newcommand{\xtagging}[1] {\ensuremath{\textrm{#1-tagging}}}
\newcommand{\xmistag} [1] {\ensuremath{\textrm{#1-mistag}}}

\newcommand{\btagged}  {\xtagged {b}}

\newcommand{\ttagged}  {\xtagged {t}}
\newcommand{\ttagging} {\xtagging{t}}
\newcommand{\tmistag}  {\xmistag {t}}

\newcommand{\ljets}  {\ensuremath{\ell+\textrm{jets}}}

\newcommand{\figref} [1] {Figure~\ref{#1}}
\newcommand{\tabref} [1]  {Table~\ref{#1}}

\begin{document}
\begin{titlepage}
\pubblock

\vfill
\Title{%
 Searches for top-antitop quark resonances
 \\[.25cm]
 at $\sqrt{s}=13\TeV$
 with the CMS detector}
\vfill
\Author{Marino Missiroli, on behalf of the CMS Collaboration} 
\Address{\institute}
\vfill
\begin{Abstract}

We present the first results on
searches for new massive resonances
decaying to a top-antitop quark ($\ttbar$) pair
with the CMS detector at the LHC
in proton-proton collisions with
a center of mass energy of $\XIIItev$.
The data set considered corresponds to
an integrated luminosity of $2.6\ifb$
recorded during the the first year of the LHC Run-2.
Searches are performed by measuring
the invariant mass distribution of the $\ttbar$ system
in semileptonic and fully-hadronic final states.
Dedicated techniques are used to identify
the decay of highly-boosted top quarks
in order to maximize the analyses' sensitivity
for $\ttbar$ resonances with a mass above the TeV scale. 
%
No significant excess is observed in the data
compared to the expected SM background and
exclusion limits are set on the cross section of
a $\ttbar$ resonance in various new physics scenarios.

\end  {Abstract}
\vfill
\begin{Presented}
$9^{th}$ International Workshop on Top Quark Physics\\
Olomouc, Czech Republic,  September 19--23, 2016
\end{Presented}
\vfill
\end{titlepage}
\def\thefootnote{\fnsymbol{footnote}}
\setcounter{footnote}{0}

The existence of new massive particles
coupling to the top quark is predicted
by numerous extensions of the Standard Model (SM).
Examples of such theories include
two-doublet Higgs models,
theories predicting additional gauge interactions,
e.g. models based on technicolor,
and models with additional space-time dimensions.
These new heavy particles could be observed
at the Large Hadron Collider (LHC) at CERN
as resonances in the invariant mass spectrum
of a $\ttbar$ pair.
%
In this document we report on
the first results of two independent searches
for $\ttbar$ resonances in $\pp$ collisions
at a center of mass energy of $\XIIItev$
with data collected by
the CMS experiment~\cite{CMS}
at the LHC
for a total integrated luminosity of $2.6\ifb$.
One search is performed on events with
one muon or electron in the final state
(semileptonic or lepton-plus-jets
analysis)~\cite{CMS-PAS-B2G-15-002},
while the other targets events
in which both top quarks decay hadronically
(all-hadronic analysis)~\cite{CMS-PAS-B2G-15-003}.


Both searches are optimized to probe
the existence of new $\ttbar$ resonances
with masses well above the TeV scale.
This requires the use of dedicated techniques
for the efficient reconstruction
of highly-boosted top quarks.
In the case of the top quark hadronic decay,
this corresponds to using a method to tag
large-radius jets comprising all the decay products
of the top quark decay and
characterized by specific substructure properties.
The two analyses make use of
the same jet $\ttagging$ algorithm,
based on two jet variables:
the groomed jet mass determined by
the softdrop algorithm, 
shown in \figref{fig:ljets__jet_Msd},
and the $N$-subjettiness ratio
$\tau_{32} \equiv \tau_{3}/\tau_{2}$~\cite{CMS-PAS-JME-15-002}.
The $\ttagging$ selection is tuned
to yield a $\tmistag$ rate of
approximately $3\%$.


The semileptonic analysis~\cite{CMS-PAS-B2G-15-002}
selects events containing
exactly one muon or electron,
at least two high-$\pt$ jets and
missing transverse energy. 
%
The lepton isolation requirement is replaced
by a selection based on the transverse momentum of
the lepton with respect to the axis of its nearest jet.
This method is found to be the most effective one
to reject non-prompt and fake leptons from multijet production
and preserve the efficiency for leptons coming from high-$\pt$ top quarks.
%
The full kinematic reconstruction of the $\ttbar$ system
is performed using the reconstructed objects in the final state.
For each event the best hypothesis is selected
based on a $\chi^{2}$ discriminator and
an upper cut on the $\chi^{2}$ value is finally applied
to reduce the contribution of non-$\ttbar$ backgrounds.
The main backgrounds in the final event sample
are given by $\Wjets$ and $\ttbar$ production.
Both backgrounds are modeled with MC simulation
and their central predictions are corrected
using data in dedicated control regions.
The final event sample is split in exclusive categories
based on the number of $\btagged$ and $\ttagged$ jets in the event.
%
\figref{fig:ljets__mttbar} shows
the $\mttbar$ distribution in the two event categories
with the highest signal efficiency.
Good agreement between data and background expectation
is found in the final $\mttbar$ spectra and
no significant excess consistent with the presence of
resonant $\ttbar$ production is observed.


The all-hadronic analysis~\cite{CMS-PAS-B2G-15-003}
selects events with a dijet topology,
requiring two large-radius high-$\pt$ jets
both passing the $\ttagging$ selection.
The main backgrounds in the final event sample
come from QCD multijet production and $\ttbar$ production.
The latter background is modeled with
a next-to-leading order MC simulation,
while the former one is determined with
a data-driven method based on
the measurement of the $\tmistag$ rate
in data control regions.
The $\mttbar$ distributions are measured
in six exclusive categories with
different signal sensitivities and background composition.
The categorization is based on
the value of the rapidity difference
between the two jets ($\Delta y_{jj}$)
and the number of $\btagged$ subjets in the events.
The $\mttbar$ distributions
in the two most sensitive categories,
given by events with $|\Delta y_{jj}|<1.0$
and at least one $\btagged$ subjet,
are shown in \figref{fig:alhad__mttbar}.
No significant deviation from the expected SM background
is observed in the mass spectra of
the reconstructed $\ttbar$ pair.



\begin{figure}[!t]
 \centering
 \includegraphics[scale=.35]{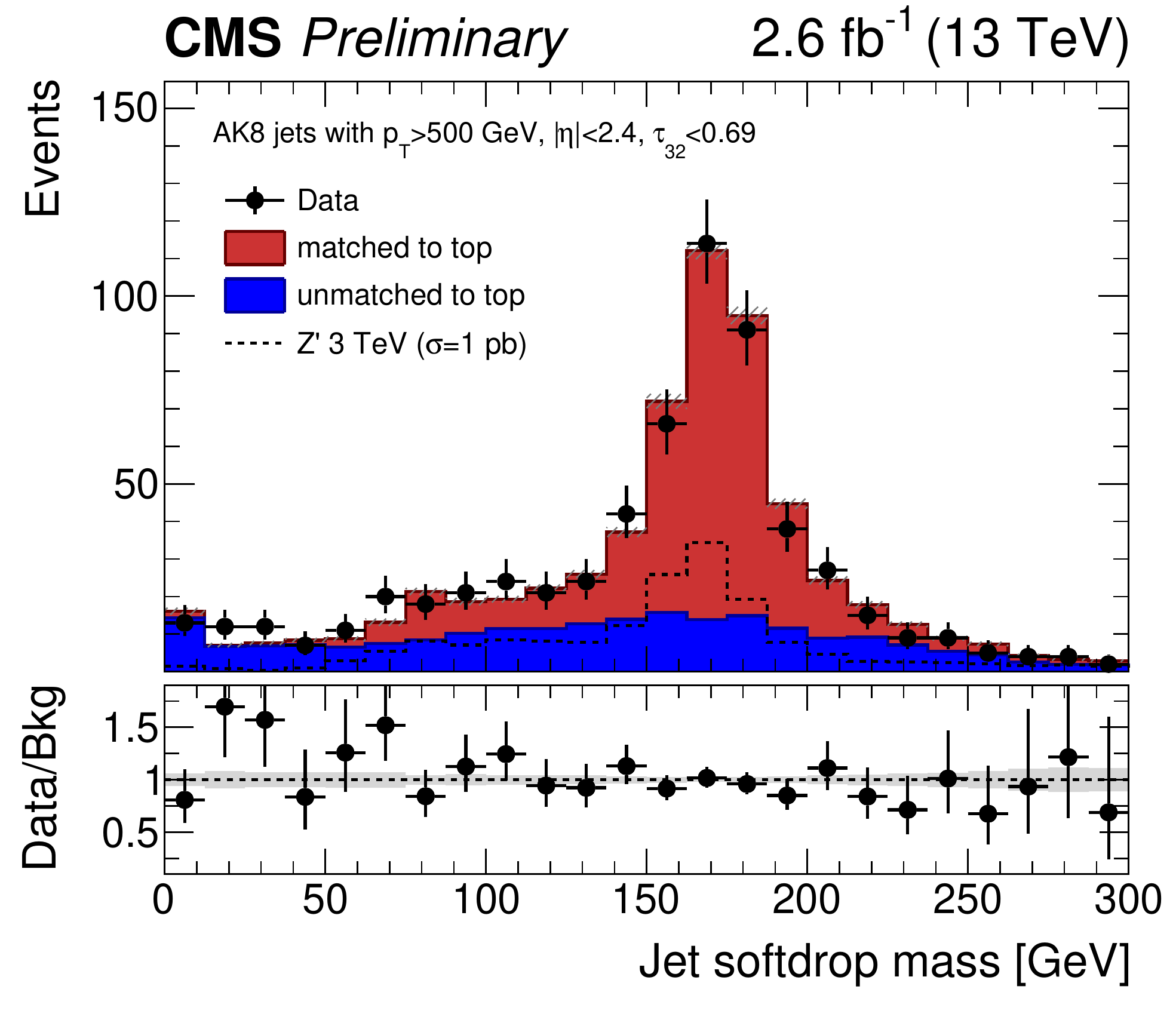}

 \vspace*{-0.50cm}

 \caption{\label{fig:ljets__jet_Msd}%
  Distribution of the softdrop mass of large-radius jets
  with $\pt > 500\GeV$, $|\eta|<2.4$ and $\tau_{32}<0.69$
  for data and simulation
  in the semileptonic
  analysis~\cite{CMS-PAS-B2G-15-002}.
  The simulation prediction is shown separately
  for jets matched (red) and not matched (blue)
  to a top quark at generator level.
  The simulation error band includes
  only the MC statistical uncertainty.
 }

\end{figure}

Since no evidence of
resonant $\ttbar$ production is observed,
the $\mttbar$ spectra are used to set
upper exclusion limits on
the cross section of a $\ttbar$ resonance with a mass up to $4\TeV$
in various new physics scenarios.
These include models with
a $\Zprime\rightarrow\ttbar$ resonance
with different width hypotheses and
a Kaluza-Klein gluon resonance
in extradimensional models.
\figref{fig:limit_xsec} shows
the cross section limits
set by the two searches
for one of these signal models.
%
These upper cross section limits are also recast
in terms of lower limits on the mass of
a $\ttbar$ resonance.
This is shown in \tabref{tab:limit_mass},
which also includes for comparison
the corresponding limits set by
the combination of $\ttbar$ searches
performed by the CMS experiment at $\VIIItev$~\cite{CMS-B2G-13-008}.
Despite the limited amount of integrated luminosity,
the lower mass limits set by these first \mbox{Run-2} analyses
are already competitive or even improved
with respect to previous searches
thanks to the increased center of
mass energy of $\XIIItev$ reached by the LHC.
In particular, the lower mass limits for large-width signals,
e.~g. $\Zprime$ resonances with a width of $10\%$,
are significantly extended.



\clearpage

\begin{figure}
 \centering

 \includegraphics[scale=.35]{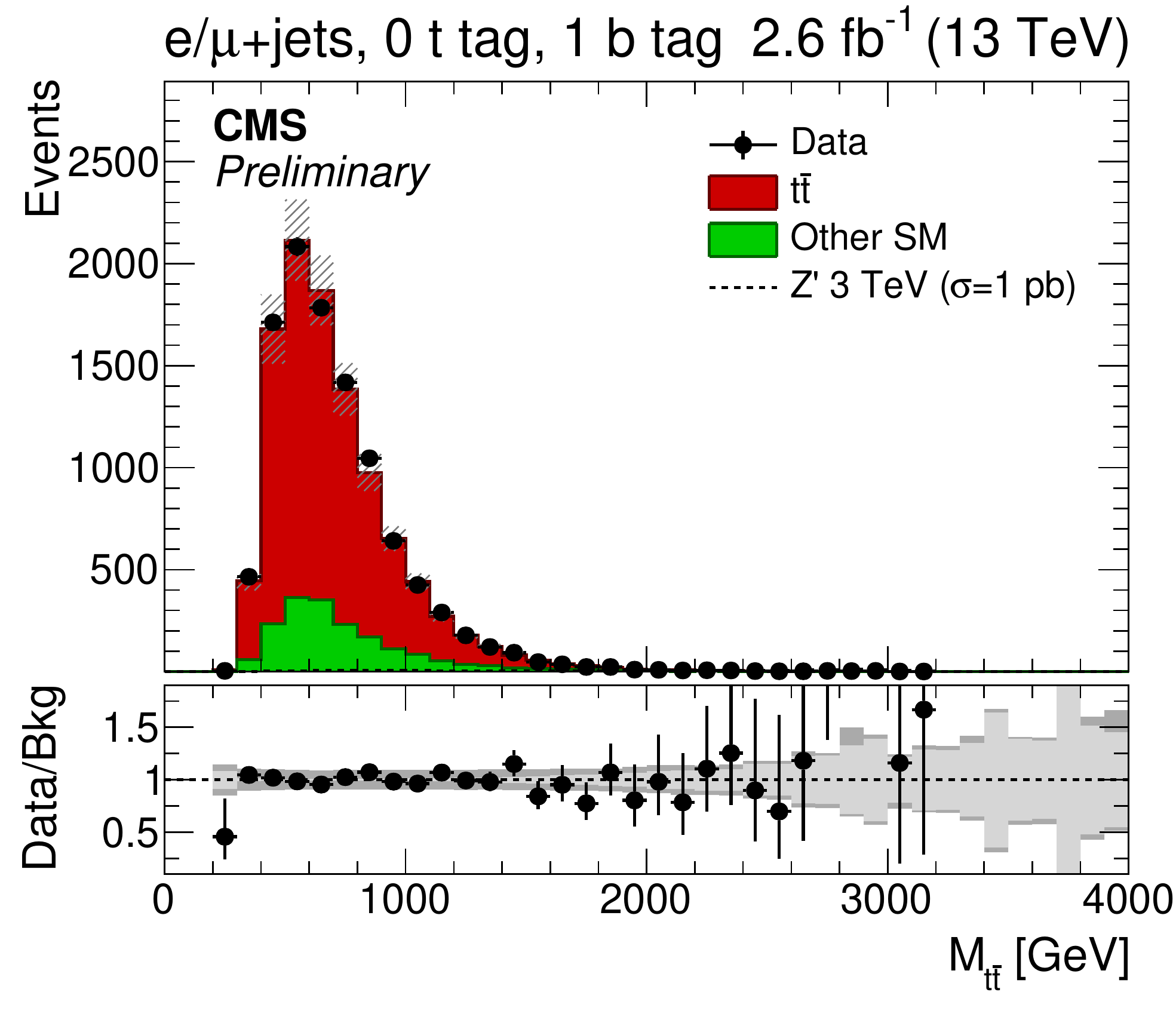}
 \hspace*{+.50cm}
 \includegraphics[scale=.35]{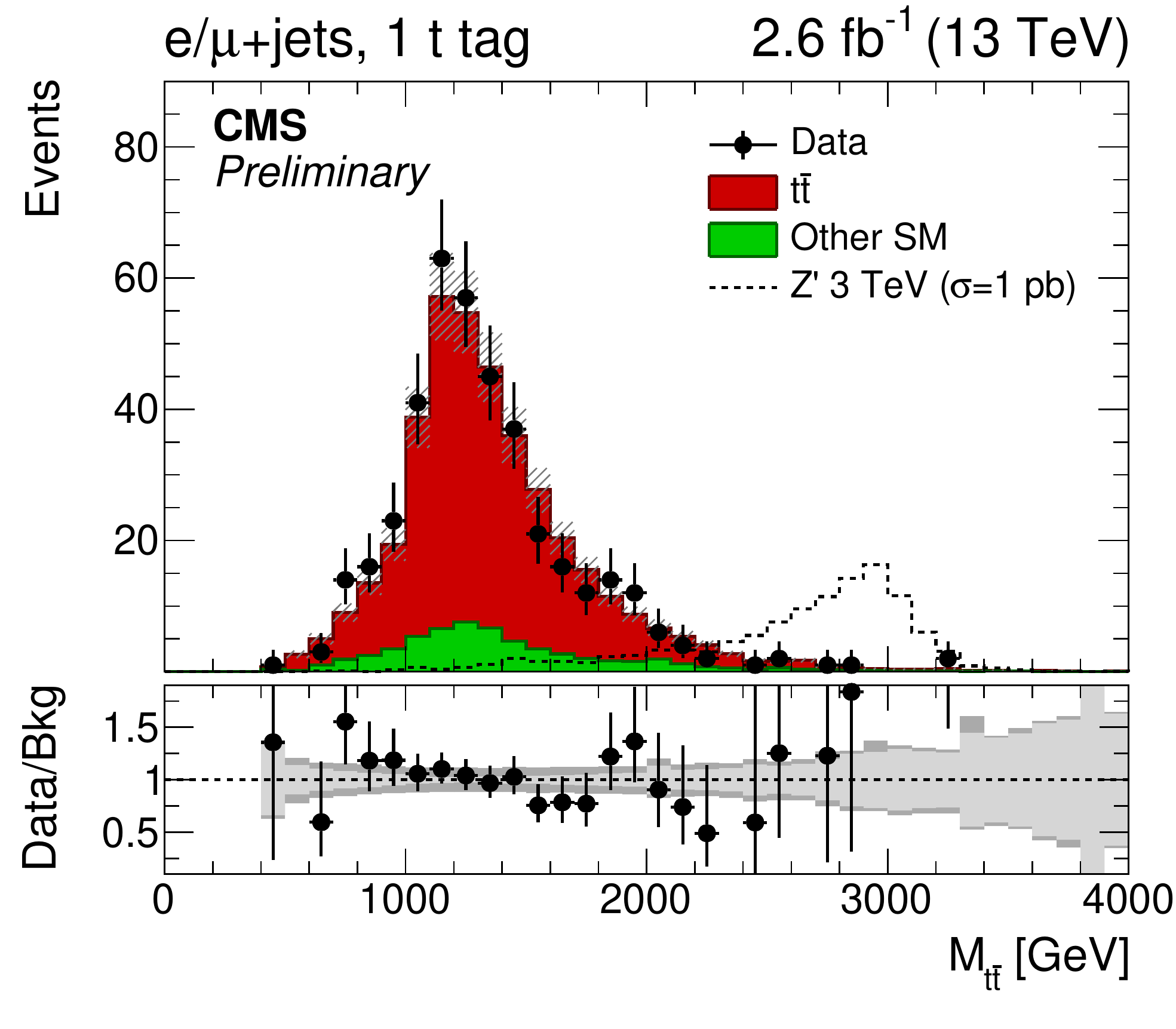}

 \vspace*{-0.25cm}

 \caption{\label{fig:ljets__mttbar}%
  Post-fit distributions of the invariant mass of
  the reconstructed $\ttbar$ pair
  in the semileptonic analysis
  for events without a $\ttagged$ jet and
  with at least one $\btagged$ jet (left)
  and for events with a $\ttagged$ jet (right).
  The signal template shown in each plot
  is normalized to a cross section of $1\pb$.
  The uncertainties associated to
  the background expectation include
  statistical and systematic
  uncertainties.
  The ratio of data to SM background
  is shown in the bottom panel of each plot,
  where the statistical (light gray) and
  total (dark gray) uncertainties
  are displayed separately~\cite{CMS-PAS-B2G-15-002}.
 }

\end{figure}


\begin{figure}
 \centering

 \includegraphics[scale=.35]{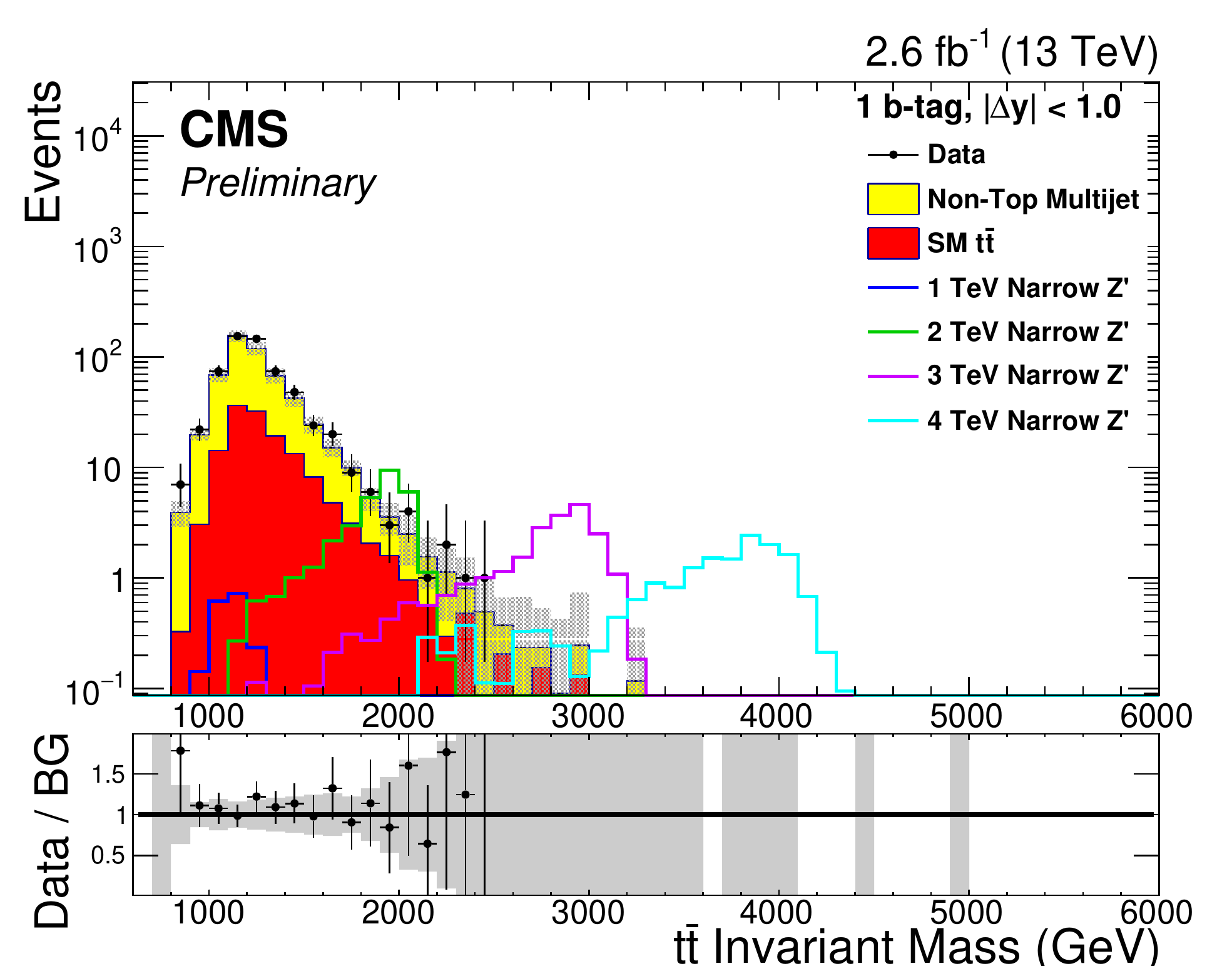}
 \hspace*{+.50cm}
 \includegraphics[scale=.35]{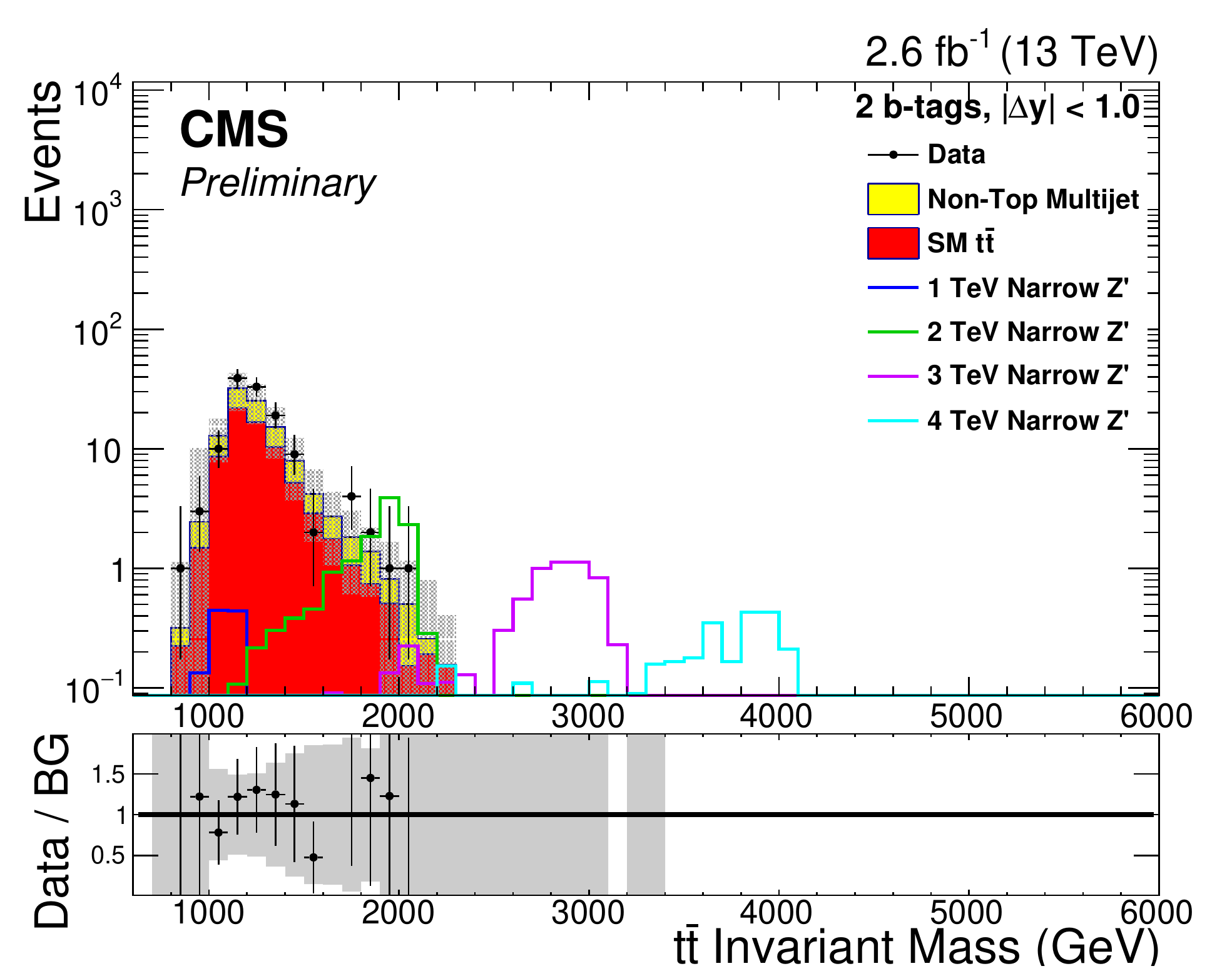}

 \vspace*{+.05cm}

 \caption{\label{fig:alhad__mttbar}%
  Post-fit distributions of the invariant mass of
  the reconstructed $\ttbar$ pair
  in the all-hadronic analysis,
  for events with $|\Delta y_{jj}|<1.0$ and
  exactly one $\btagged$ subjet (left) or
  at least two $\btagged$ subjets (right).
  Each signal template
  is normalized to a cross section of $1\pb$.
  The uncertainties associated to
  the background expectation include
  statistical and systematic
  uncertainties.
  The ratio of data to SM background
  is shown in the bottom panel of each
  plot~\cite{CMS-PAS-B2G-15-003}. 
 }

\end{figure}


\clearpage

\begin{figure}
 \centering

 \vspace*{-0.50cm}

 \raisebox{+0.1\height}{\includegraphics[scale=.335]{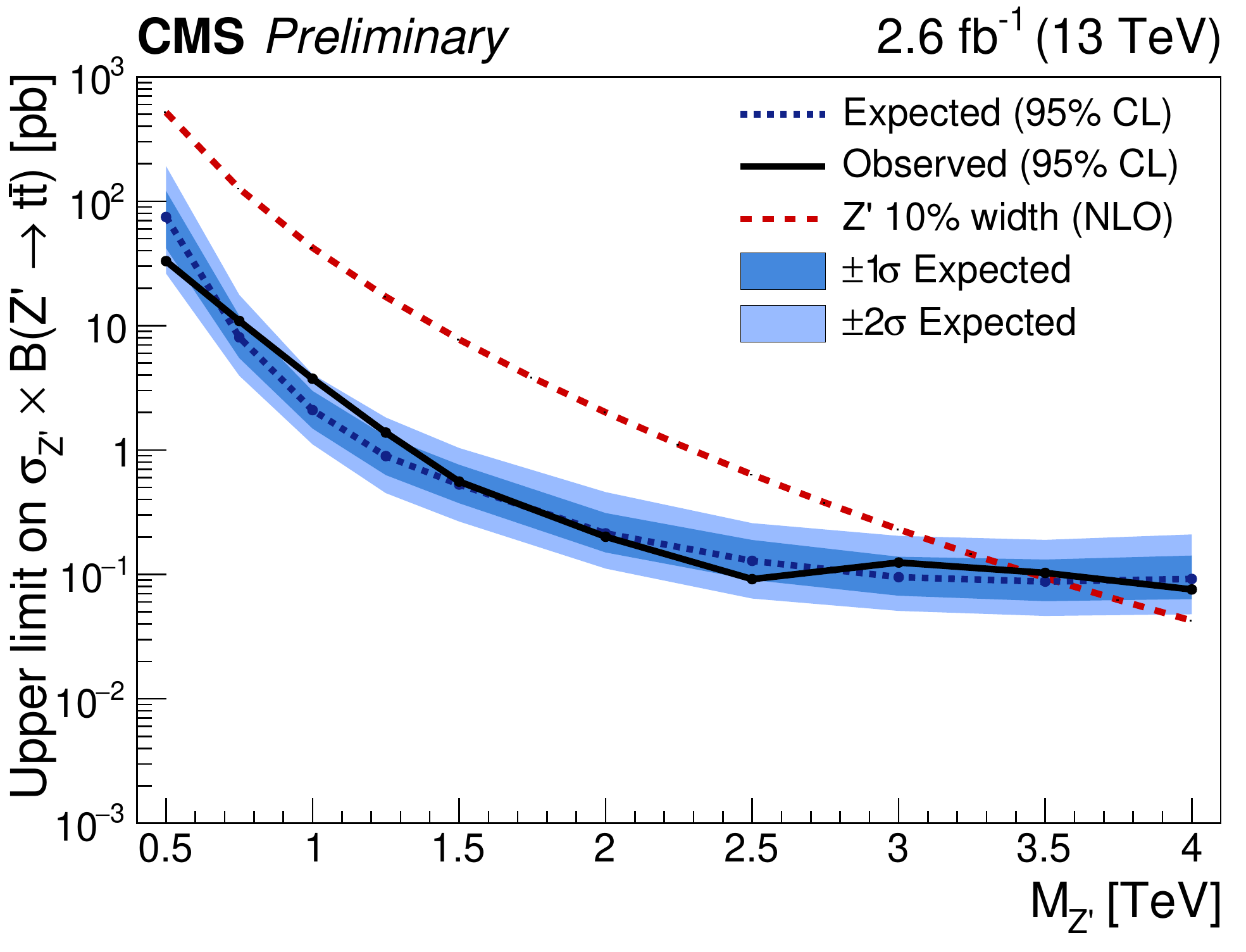}}
 \hspace*{+0.50cm}
 \raisebox{+0.0\height}{\includegraphics[scale=.325]{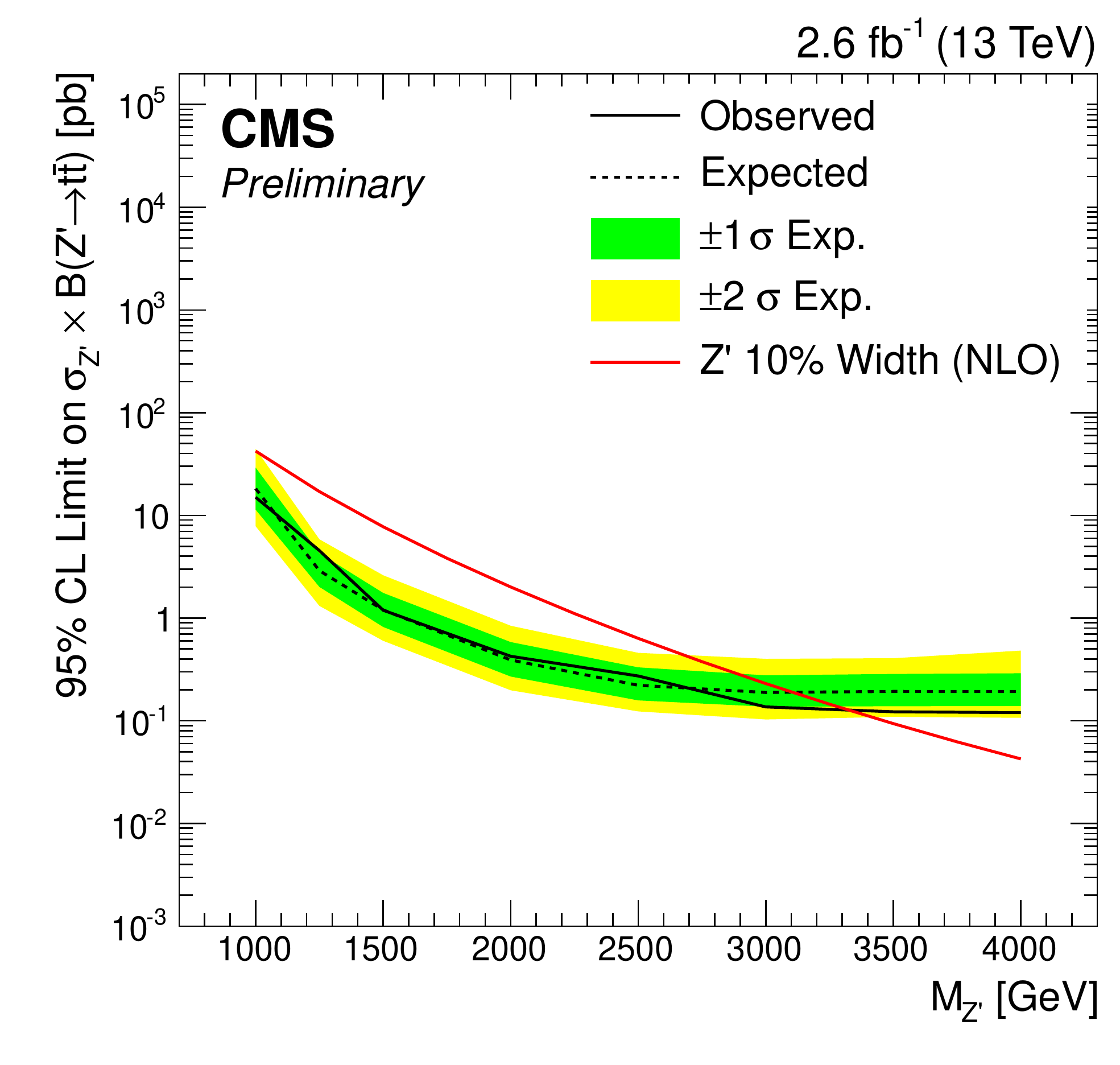}}

 \vspace*{-0.50cm}

 \caption{\label{fig:limit_xsec}%
  Expected and observed $95\%$ CL upper limits on
  the production cross section times branching ratio
  for a ${\Zprime\rightarrow\ttbar}$ resonance
  with a relative decay width
  (${\Gamma_{\Zprime}/M_{\Zprime}}$)
  of $10\%$.
  Limits are shown for
  the semileptonic analysis~\cite{CMS-PAS-B2G-15-002} (left) and
  the all-hadronic analysis~\cite{CMS-PAS-B2G-15-003} (right)
  as a function of the signal mass hypothesis.
 }

\end{figure}

\vspace*{-1.25cm}

\begin{table}[htbp]
 \centering

 \caption{\label{tab:limit_mass}%
  Observed mass exclusion range
  for the $X\rightarrow\ttbar$ signal models
  considered in the semileptonic and all-hadronic analyses.
  The corresponding values from the combination of
  $\ttbar$ searches in CMS at $\VIIItev$~\cite{CMS-B2G-13-008}
  are also shown.
 }

 \vspace*{.35cm}

 \scalebox{.88}{
  \begin{tabular}{|lr|c|c|c|c|}
   \multicolumn{6}{l}{\textbf{Observed mass exclusion region [TeV]}}                                     \\[0.1cm] \hline \hline \rule{-2pt}{12pt}
    channel      & & $\Zprime$  ($ 1\%$ width)
                   & $\Zprime$  ($10\%$ width)
                   & $\Zprime$  ($30\%$ width)
                   & KK gluon                                                                            \\[0.1cm] \hline        \rule{-2pt}{12pt}
    $\ljets$     & ($13\TeV$)~\cite{CMS-PAS-B2G-15-002} & $0.6-2.3$ & $0.5-3.4$ & $1.0-4.0$  & $0.5-2.9$ \\[0.1cm]               \rule{-2pt}{12pt}
    all-hadronic & ($13\TeV$)~\cite{CMS-PAS-B2G-15-003} & $1.4-1.6$ & $1.0-3.3$ & $1.0-3.8$  & $1.0-2.4$ \\[0.1cm] \hline        \rule{-2pt}{12pt}
    combination  &  ($8\TeV$)~\cite{CMS-B2G-13-008}     & $0.5-2.4$ & $0.5-2.9$ & not tested & $0.5-2.8$ \\[0.1cm] \hline
  \end{tabular}
 }

\end{table}

\vspace*{-0.40cm}

\Acknowledgements

The financial support of
the Spanish Ministerio de Econom\'{i}a y Competitividad,
under project number FPA2014-53938-C3-3-R,
is acknowledged.

\end{document}